\documentclass[twocolumn]{aastex631}

\usepackage{amsmath}
\usepackage{multirow}
\usepackage{tabularx}
\usepackage{diagbox}
\usepackage{rotating}

\newcommand{\xs}{{\tt XS}}

\newcommand{\kms}{km\,s$^{-1}$}

\newcommand{\PAdisk}{$\phi_\mathrm{disk}^\prime$}

\newcommand{\sn}{S$/$N}
\newcommand{\hi}{\ion{H}{1}}

\newcommand{\ha}{H$\alpha$}

\newcommand{\comm}[1]{}

\received{June 1, 2019}
\revised{January 10, 2019}
\submitjournal{AJ}

\shorttitle{NC motions in MaNGA}
\shortauthors{Lopez-Coba et al.}
\begin{document}

\title{On the role of non-circular motions in MaNGA galaxies I: global properties}

\correspondingauthor{C. Lopez-Coba}
\email{calopez@asiaa.sinica.edu.tw}

\author[0000-0003-1045-0702]{Carlos L\'opez-Cob\'a}
\affiliation{Institute of Astronomy and Astrophysics, Academia Sinica, No. 1, Section 4, Roosevelt Road, Taipei 10617, Taiwan}

\author{Lihwai\,Lin}
\affiliation{Institute of Astronomy and Astrophysics, Academia Sinica, No. 1, Section 4, Roosevelt Road, Taipei 10617, Taiwan}

\author{Justus\,Neumann}
\affiliation{Max-Planck-Institut f\"{u}r Astronomie, K\"{o}nigstuhl 17, D-69117 Heidelberg, Germany}

\author{Matthew\,A.\,Bershady}
\affiliation{University of Wisconsin - Madison, Department of Astronomy, 475 N. Charter Street, Madison, WI 53706-1582, USA}



\begin{abstract}
Non-circular (NC) motions represent the imprints of non-axisymmetric structures in galaxies, providing opportunities to study the physical properties of gas departing from circular rotation. In this work we have conducted a systematic study of the non-circular motions in a sample of 1624 gas-rich disk galaxies from the MaNGA MPL-11. By using the \ha~velocity as a tracer of the disk rotation, we find indications that the amplitude of the non-circular motions is related to the stellar mass, with the low mass and late-type galaxies the most affected. In our sample, we find ratios of non-circular to circular rotation ranging from 5\% to  20\%.
By implementing harmonic models to include NC motions associated with spiral arms and stellar bars, we find that the rotational curves traced with \ha~are barely affected by the NC induced by these structures. Consequently, in our sample, we do not find evidence that NC motions contribute to the scatter of the stellar Tully-Fisher relation. Our results suggest that non-circular motions might have a more localized effect in galaxies rather than a global one.
\end{abstract}



\section{Introduction} \label{sec:intro}

The existence of kinematic asymmetries or non-circular (NC) motions in disk-galaxies is known since the first modelings of the line-of-sight (LoS) velocity in nearby galaxies \citep[e.g.,][]{Warner1973,Bosma1978,Begeman1987PhD,Begeman1989}. These studies followed by many others showed that in presence of non-axisymmetric structures, kinematic patterns appeared in the residual velocity maps, indicating the presence of sources of non-circular rotation \citep[e.g.,][]{Trachternach2008, Andersen2013, Oman2019, Lang2020};
hence, they appear to be ubiquitous in galaxies, regardless of their stellar mass and morphological type.

The most common sources of non-circular rotation at sub-kpc scales are due to non-axisymmetric structures, including spiral arms \citep{Davies2009,vandeVen2010}, oval structures such as bulges and stellar bars \citep{Pence1984,Holmes2015, LopezCoba2022},  warps \citep{Wong2004} among others. Although the presence of large scale galactic winds can contribute significantly to the NC motions observed in galaxies \citep{clc2018, AMUSING++, Lacerda2020}.

Despite knowing its existence, it is uncertain how  non-circular motions affect to the local and global processes in galaxies. Often, non-circular motions on the disk plane are interpreted as radial, or streaming flows that redistribute metals and gas through the disk \citep[e.g.,][]{Tremonti2004}.
Due to loss of angular momentum these flows can reach the inner regions of a galaxy and provide fuel to the central active galactic nuclei, AGN \citep[][]{Combes2001}, or trigger nuclear star formation \citep[][]{Davies2009}. Eventually, the infall of gas to the center will trigger the formation of galactic winds. This process continues until the gas is consumed and the galaxy quenches due to the lack of fuel.

In the above scenario, non-axisymmetric structures like spiral arms or stellar bars act as rivers through which the gas flows; therefore, characterizing the kinematics perturbations induced by these structures is of crucial importance to understand the role they play in the star formation processes in galaxies.

In practice, however, associating non-circular motions to radial in/out-flows requires {\it ad-hoc} hydrodynamic modeling \citep[e.g.,][]{Davies2009,Davies2014}. Although this approach provides a more detailed description of the phenomenon, it makes it difficult to systematize in large galaxy samples.
On the other hand, mild non-circular motions linked to perturbations to the gravitational potential have been identified in the velocity field of disk galaxies \citep[e.g.,][]{Schoenmakers1997} and in residual velocity maps \citep{Erroz-Ferrer2015}.

It is still under debate whether there is a relationship between non-axisymmetrically induced non-circular motions and the local properties in galaxies \citep{Erroz-Ferrer2015, LopezCoba2022}. This is mainly due to our inability to properly characterize them.
In fact, there is no unambiguous method for studying the contribution of non-circular motions in galaxies, mostly because this implies being able to decouple the non-circular contribution from the circular rotation, which often adopts certain assumptions.

Thanks to large spectroscopic galaxy surveys like the Mapping Nearby Galaxies at Apache Point Observatory \citep[MaNGA,][]{ManGA}, it is possible to gain  insights into the role that NC motions play in galaxies. The large spatial coverage of MaNGA, spanning between 1.5--2.5 effective radii, along with its moderate spectral resolution, makes it ideal for this purpose.

This paper is organized as follows. In section~2 we describe the data and the sample selection; the analysis and kinematic models adopted are described in section~3; in section~4 we present the results and finally in section~5 the conclusions. Throughout this paper we adopted a $\Lambda$CDM cosmology with $H_0 = 71$~\kms~Mpc$^{-1}$, $\Omega_m = 0.3$ and $\Omega_{\Lambda} = 0.7$.

\section{Data} \label{sec:data}
For this study, we use the MaNGA Product Launch-11 (MPL-11) which consists of $\sim10000$ galaxies from the Local Universe.
MaNGA is an integral field spectroscopic galaxy survey that observed nearby galaxies at $z\lesssim 0.1$, under the following instrumental configurations: an average spectral resolution of $R\sim2000$ and average spatial resolution at the full-width at half maximum, $\mathrm{FWHM}\sim 2.5\arcsec$ \citep[e.g.,][]{Law2016}.

In particular in this study, we used the dataproducts analyzed with the {\sc pypipe3d} pipeline \citep[e.g.,][]{Lacerda2022} and described in \cite{Sebastian2022}. The dataproducts from this pipeline contain two-dimensional information about the ionized gas content and the stellar population properties of each of the analyzed MaNGA galaxies. In this work, we will extensively use the ionized gas properties (fluxes and their corresponding velocity maps), and the stellar properties, such as equivalent widths and stellar mass density maps. We refer to \cite{Sebastian2022}  for a thorough description of the stellar population analysis of the MaNGA datacubes.

Additionally, in this work we used the sky projection angles,  disk inclination ($i$) and position angle ($\phi_{disk}^{\prime}$), as well as the effective radius ($r_e$), all of which obtained from the NASA-Sloan Atlas catalogue \citep[NSA,][]{nsa}.

\subsection{Sample selection}
We are interested in analyzing the gas kinematics in disk galaxies, for which we focus on star-forming (SF) galaxies with plenty of ionized gas.
As a first filter, we selected objects from MaNGA that show an equivalent width of \ha~at the effective radius larger than 3\AA~\citep{Lacerda2018, Sebastian_ARAA}. This criterion should primarily exclude early-type galaxies, whose kinematics are generally dominated by random motions and therefore are not suitable for kinematic modelling.

Then, from these filtered galaxies we select those with the major coverage of their optical extension, following the criteria below. (i) galaxy redshifts must lie between $0.005<z<0.055$, this ensures maximum spatial scales of $1$~kpc/\arcsec; (ii) objects should be observed by more than 91-fiber integral field units (IFU) to maximize the number of independent pixels within the FoV (field-of-view),  which is important for the rotation curve estimation; (iii) the effective radius ($r_e$) should be larger than the FWHM resolution; (iv) the ratio between the cube FoV diagonal radius and $r_e$ should be less than or equal to 3 to avoid selecting objects with apparent small sizes relative to their FoV; (v) disk inclinations should lie within $30^{\circ}<i<70^{\circ}$, where $i$ is the angle
formed between a perpendicular vector to the disk-plane and the LoS direction. On one hand, the circular rotation in nearly face on galaxies is compromised with the inclination angle ($\sin i \sim i $ for $i\sim0^{\circ}$), resulting in large uncertainties in the estimation of the rotation curve. Similarly, recovering the true circular velocity in highly inclined systems becomes challenging due to their complex line-of-sight velocity distribution, often necessitating dedicated modeling \citep[][]{Kregel2004}.

\begin{figure*}[t]
%\centering\includegraphics[width = 0.8\textwidth]{/media/carlos/ofelia/MaNGA_non_circ/figures/sample}
\centering\includegraphics[width = 0.8\textwidth]{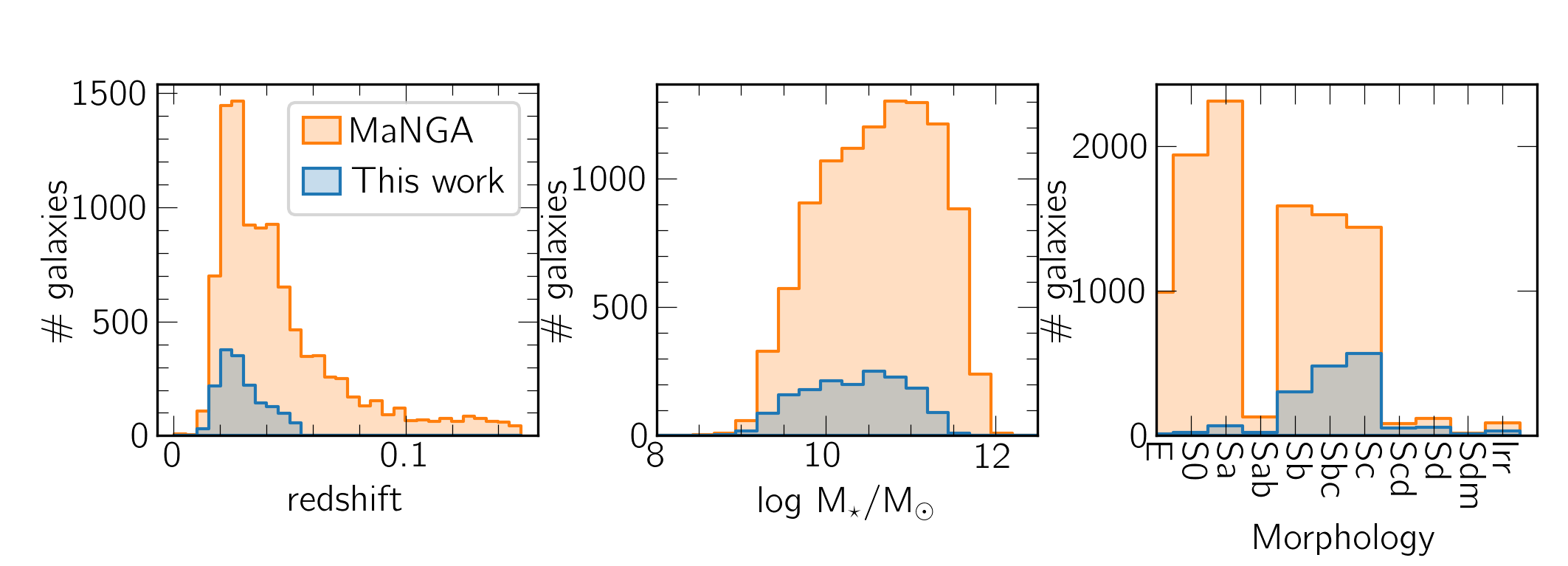}
\caption{{\it From left to right:} distribution of redshift, stellar mass and morphology of the kinematic sample (blue) compared to the whole MaNGA galaxies (orange). The p-value ($\ll 0.05$) from a KS test indicates that our kinematic sample does not represent the entire MaNGA distribution; instead it is biased towards late-type spirals.
}
\label{fig:sample}
\end{figure*}

The implementation of the aforementioned criteria resulted in 1624 candidate galaxies, defining our sub-sample for kinematic study, hereafter referred to as the kinematic sample. Figure~\ref{fig:sample} shows the distribution of redshift, stellar mass, and morphological type for our kinematic sample compared to the entire MaNGA galaxies. A KS test on these distributions reveals that our kinematic sample is not representative of the entire MaNGA sample but is biased towards disk galaxies, particularly late-type spirals.

\section{Analysis}
\subsection{Kinematic models}
The kinematics analysis was performed with {\tt XookSuut}~\citep[\xs,][]{XookSuut}. \xs~is a Bayesian code designed to model circular and non--circular motions on velocity maps. This code creates an interpolated model over concentric rings adopting the flat disk approximation, that is, the ellipticity and the position angle are assumed constant throughout the galaxy. The latter criterion will be crucial for our analysis since, as we will see later, it favors the appearance of non-circular motions induced by non-axisymmetric structures, instead of vanishing them if variations in the position angle or inclination were allowed in the model.

In this study, we will focus on analyzing the \ha~velocity as a tracer of the circular rotation. Although the stellar velocity is a better tracer of the potential and thus of circular rotation, its spatial resolution is affected by the pixel co-adding imposed to increase the \sn~(signal-to-noise) of the continuum, which is necessary for recovering the stellar population properties \citep[e.g.,][]{Lacerda2022,Sebastian2022}.

The effects of beam-smearing on the velocity maps were corrected by deconvolving the maps following the method described in \citet{Chung2021}, using a circular Gaussian point-spread-function PSF with a $2.5\arcsec$/FWHM.

\xs~allows fitting multiple non-circular rotation models, including a general harmonic decomposition of the LoS-velocity.
In this work, we will adopt two different kinematic models to describe the MaNGA velocity fields. The circular model is described by the usual expression:
\begin{equation}
 \label{Eq:circular}
 V_\mathrm{circ}(r, \theta) = V_\mathrm{sys} + \sin i ~V_t(r)\cos \theta,
\end{equation}
where $V_\mathrm{sys}$ is the constant systemic velocity; $V_t$ is the tangential velocity often called circular rotation, which is a function of the galactocentric distance $r$;  $i$ is the disk inclination; and $\theta$ is the azimuthal angle in the galaxy plane, measured from the major-axis, and it depends on projection angles and kinematic centre.

Our second model takes into account the presence of non--circular motions through a harmonic expansion of the line-of-sight velocities. This method assumes that non--circular motions lie in the plane of the disk, and their amplitudes are small compared to the mean azimuthal velocity, denoted as $V_t$. The formalism behind it was developed by \cite{Schoenmakers1997} and it is based on the assumption that non-circular motions can be described as slight perturbations of circular orbits, transforming them into slightly elliptical closed orbits \citep[e.g.,][]{Franx1994}.% therefore, only mild non--circular motions could be described with this approach.

Instead of adopting an arbitrary expansion of the LoS velocity as often assumed \citep[e.g.,][]{kinemetry}, we adopt a model that
accounts for perturbations induced by two-armed spiral arms and  stellar bars as in \citet{Schoenmakers1997}. This model is described by the expression:
\begin{equation}
 \begin{aligned}
 \label{Eq:m=3}
 V_\mathrm{los} =  & c_0 + \sin i \big[ c_1(r)\cos \theta  +  c_3(r)\cos 3\theta  \\
                    & + s_1(r)\sin \theta  + s_3(r)\sin 3\theta \big],
\end{aligned}
\end{equation}
%
%\begin{multline}
% \label{Eq:m=3}
% V_\mathrm{los} =  c_0 + \sin i \big[ c_1(r)\cos \theta  +  c_3(r)\cos 3\theta  \\ + s_1(r)\sin \theta  + s_3(r)\sin 3\theta   \big],
%\end{multline}
%
where $s(r)$ and $c(r)$ represent the harmonic coefficients and $c_0$ is the constant systemic velocity.

Within the MaNGA sample, two-armed spirals are the most common type of spirals, and the majority of them are star-forming galaxies \citep[e.g.,][]{Hart2016} exhibiting abundant emission of \ha~through their disks. Therefore, applying Eq.~\ref{Eq:m=3} to our MaNGA sub-sample will provide us a general idea of the ``average strength'' of these coefficients.

\begin{figure*}[t]
\centering\includegraphics[width = 0.8\textwidth]{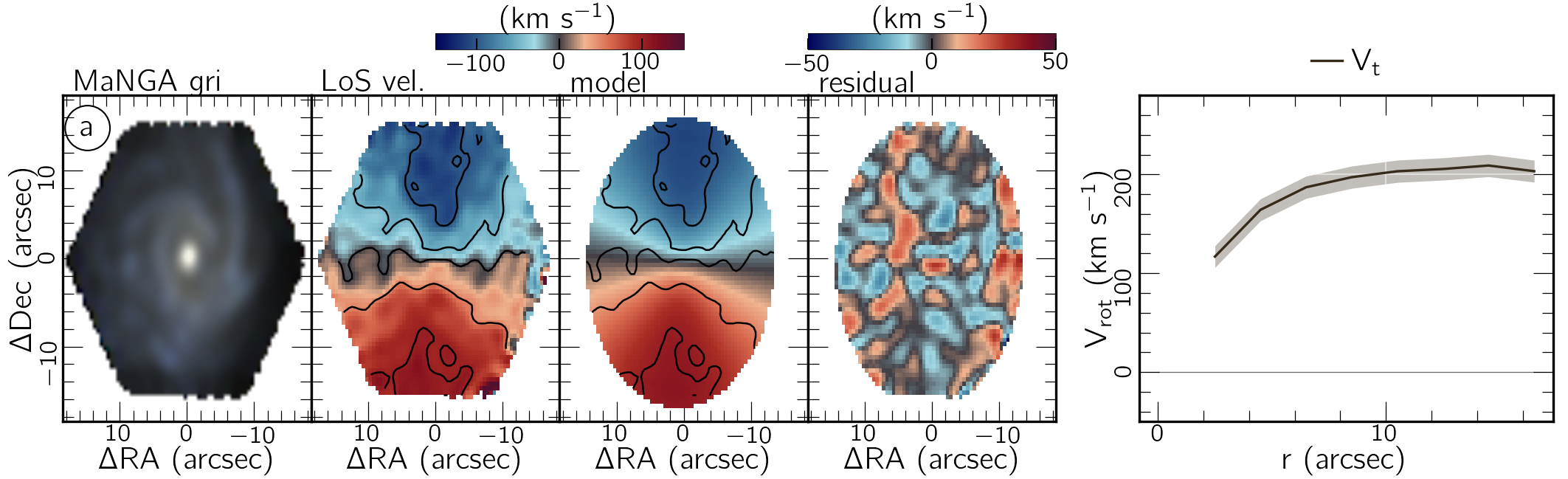}
\centering\includegraphics[width = 0.8\textwidth]{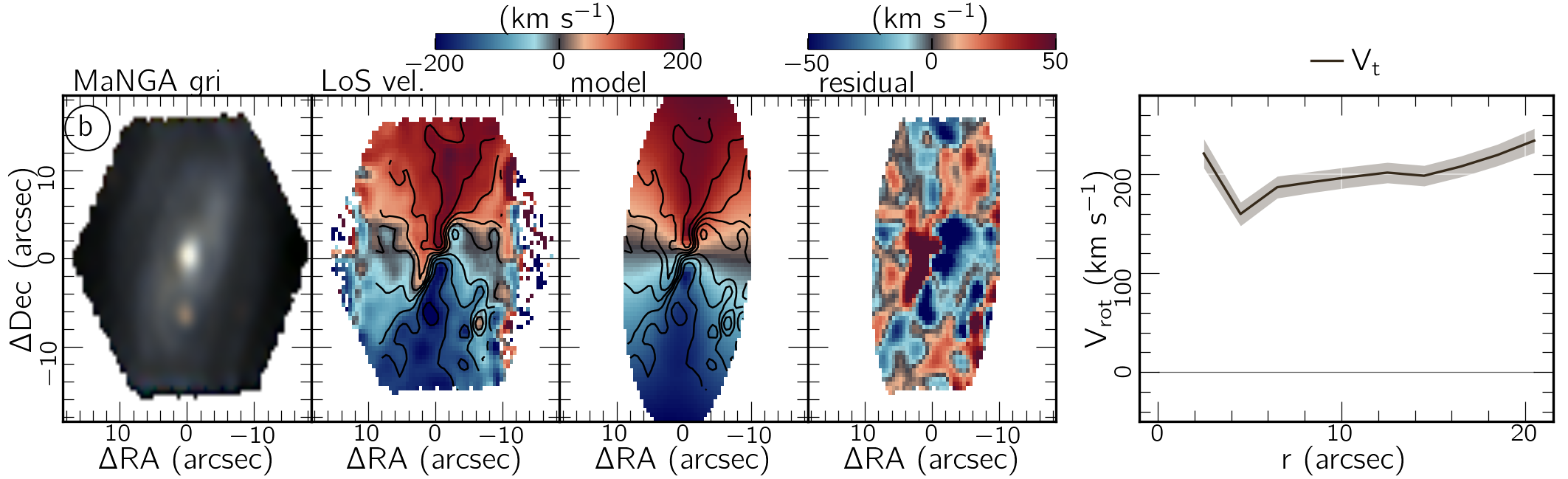}
\centering\includegraphics[width = 0.8\textwidth]{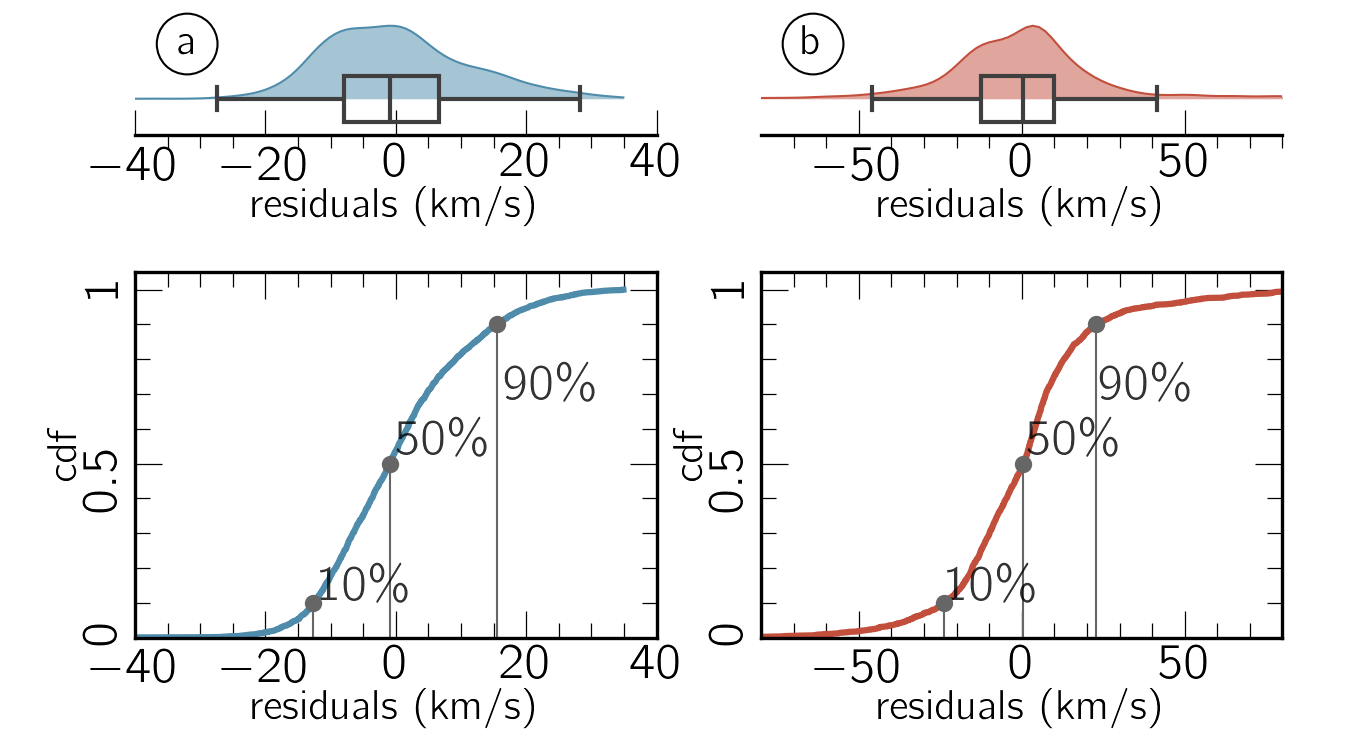}
\caption{ {  \it Top figures:}  Modeling of the \ha~velocity field adopting circular rotation only for the manga object 8147-12703 (tagged with the letter $a$), and for  10842-12704  (tagged with $b$). {\it From left to right:} $gri$ colour composite image with filters extracted from the datacube; \ha~velocity map; best circular rotation model from \xs; residual map obtained from subtracting the best kinematic model to the \ha~velocity; rotational curve. {  \it Bottom figures:}
Half-violin plot and box plot of the residual velocities of objects $a$ (blue) and $b$ (red); bottom plots show the corresponding cumulative distribution function of residuals, with vertical lines showing the 10th, 50th, and 90th percentiles.
}
\label{fig:circular}
\end{figure*}

\begin{figure*}[t]
%\centering\includegraphics[width = 0.7\textwidth]{f1}
\centering\includegraphics[width = 0.8\textwidth]{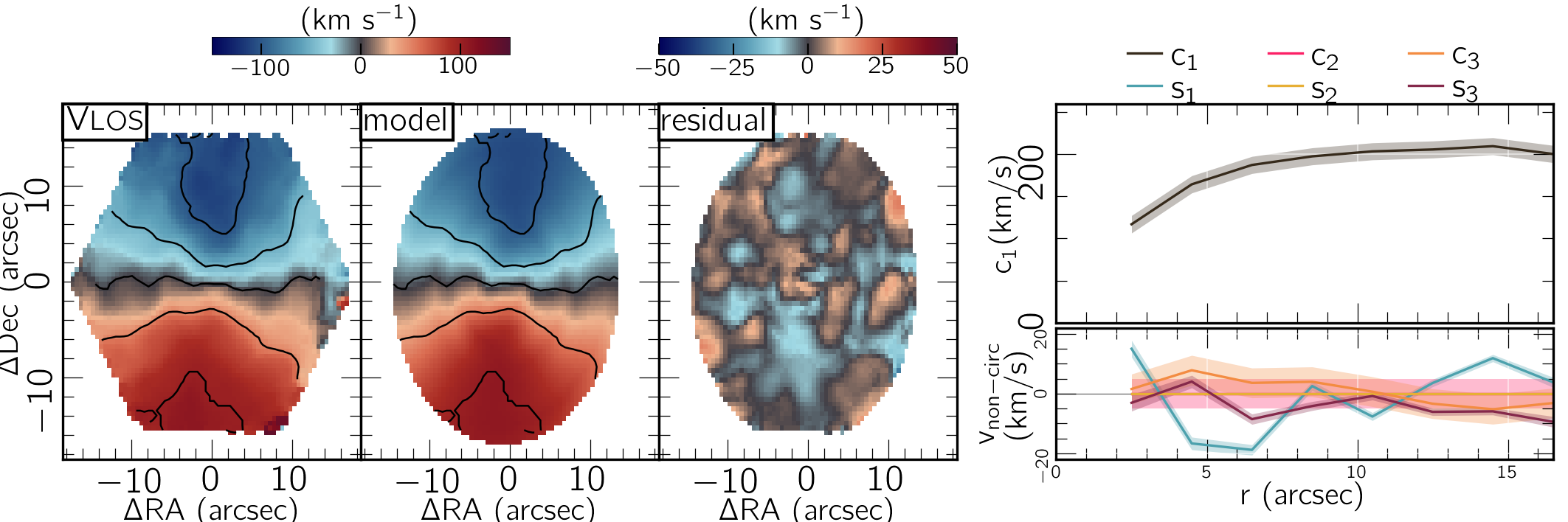}
\centering\includegraphics[width = 0.8\textwidth]{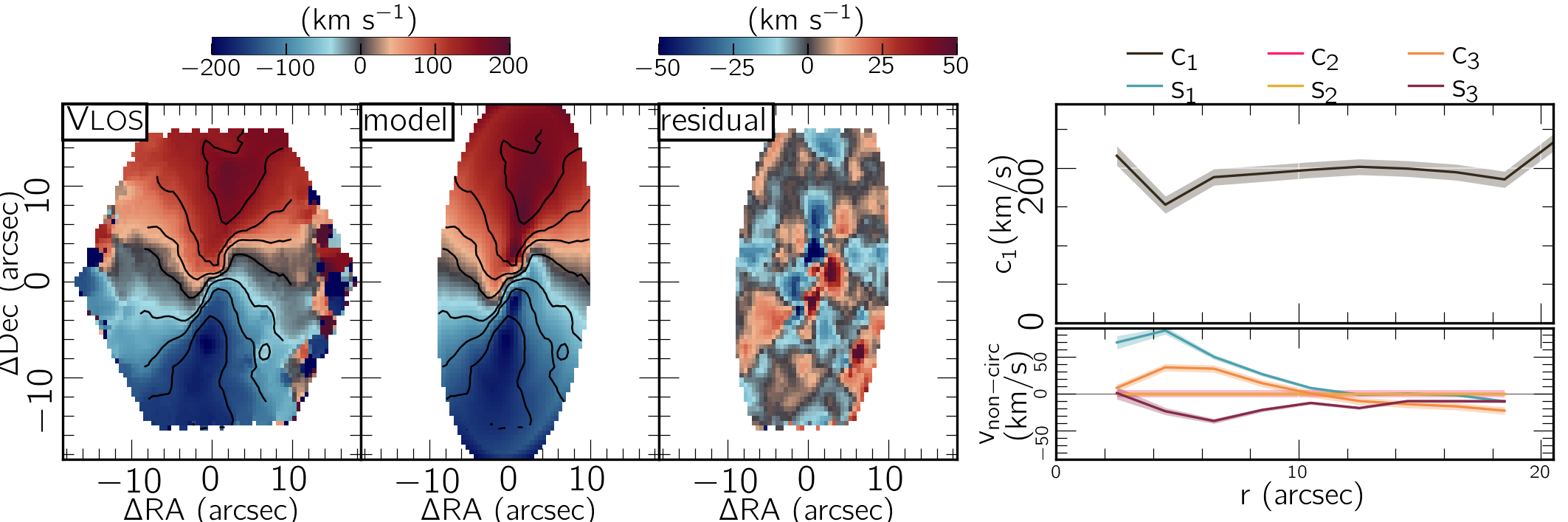}
\caption{ Similar as Figure~\ref{fig:circular}, but for the harmonic decomposition model. Note that  $s_2(r) = c_2(r) = 0~\forall~r$ as required by the model from equation.~\ref{Eq:m=3}.}
\label{fig:harmonic}
\end{figure*}

\subsubsection{Showcase examples}
Following, we built circular and non-circular rotation models for our sub-sample of MaNGA galaxies.

\xs~requires initial values for the disk position angle (\PAdisk), disk inclination $i$, kinematic centre ($x_c,~y_c$) as well as the spacing between rings. In addition, we used the velocity error maps for excluding spatial pixels (spaxels) with low signal-to-noise in \ha~that could affect the final models, specifically we removed those spaxels where errors are larger than 15~\kms.
%.
We used the NSA projection angles as the starting positions for the disk geometry, and allowed \xs~to find the best values for the gaseous disk. 
The spacing between rings was chosen to be $2.5\arcsec$, which corresponds to the average FWHM spatial resolution of the data. Additionally, the first ring was placed at a $2.5\arcsec$ distance from the kinematic centre to avoid sampling below the FWHM resolution. For the circular rotation models ({\it i.e.}, Eq.~\ref{Eq:circular}), we adopted Bayesian sampling methods to derive the velocities and the geometric parameters describing the gaseous disk (namely \PAdisk, $i$, $x_c$, $y_c$).

On the other hand, for the harmonic models we adopt a slightly different approach.
Incorrect estimation of the disk geometry could induce the appearance of artificial residual patterns \citep[e.g.,][]{Schoenmakers1997, kinemetry}, which could then be captured by the harmonic terms. To minimize these effects, we fixed the disk geometry to those values obtained from the circular rotation models. This ensures that \xs~will only estimate the
kinematic parameters ($c_k(r)$ and $s_k(r)$). If the projection angles were allowed to have a radial dependence, the final model would result in incorrect estimations of the true projection angles, but these would be compensated by variations in the amplitude of the harmonic coefficients.
Finally, we constrained the radial extension of the harmonic coefficients  up to $2r_e$.

Figure~\ref{fig:circular} shows the implementation of the circular rotation model
applied on two different objects, namely manga-8147-12703 and 10842-12704, while figure~\ref{fig:harmonic} shows results for the harmonic model.
These two objects exhibit clear non-axisymmetric structures such as spiral arms and stellar bar, while their corresponding velocity maps are presumably affected by their presence.
The first object in the former figure shows a quasi-symmetric velocity field typically from disk galaxies, with possible presence of stream motions. Multiple wiggles are observed along the minor axis probably caused by the spiral arms.
From the residual velocity map we observe different velocity patterns most likely associated to perturbations induced by the spiral arms.
On the other hand, the second object shows a clear bar in the MaNGA {\it gri} continuum image, while the velocity field shows signs of stream motions in the inner region most probability induced by the stellar bar. This kind of distortion is common from barred galaxies \citep[e.g.,][]{Holmes2015, LopezCoba2022}, and is usually described by suitable kinematic models that take into account the bar dynamics \citep[e.g.,][]{Spekkens2007}.

\begin{figure}
\centering
\includegraphics[]{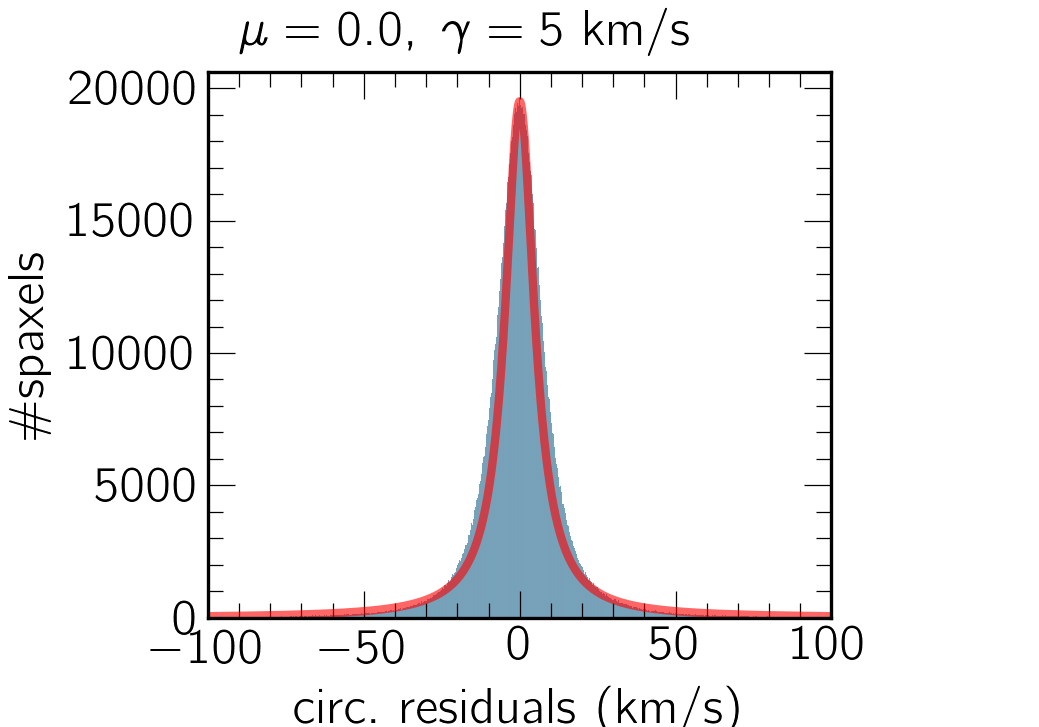}
\caption{Distribution of circular rotation residual velocities in our kinematic sample. The distribution is characterized by a Cauchy distribution, shown in red, centered in 0 with a  5 \kms~scale. }
\label{fig:hist_circ}
\end{figure}

In the harmonic models in figure~\ref{fig:harmonic}, we note that the addition of the non-circular terms (namely $s_1$, $s_3$ and $c_3$) capture small kinematic patterns that the circular model failed to reproduce. For instance, the change in orientation of the minor axis ($\sim 0$\,\kms~iso-velocity contour) is well reproduced in both galaxies.
While variations in the position angle and inclination can produce this pattern in the velocity field, it is not as frequent to observe warps within the optical extent of galaxies as is observed in \hi~disks \citep[e.g.,][]{Kamphuis2015}. Thus, the presence of streaming flows in the \ha~disks is a more likely interpretation. Moreover, our kinematic analysis favors {\it per\,ce} the appearance of non-circular motions due to non-axisymmetric structures, and random motions, rather than those due to projection effects (namely  induced by variations in $i(r), \phi_{disk}^{\prime}(r)$, $x_c(r)$, $y_c(r)$).
We are aware, however, that systematizing the analysis in this way might result in modeling galaxies where this radial dependence on the projection angles is indeed observed. A separate analysis should be performed in such objects, which is out of the scope in this work.

As observed from the residual maps in Figure~\ref{fig:harmonic}, the harmonic models still leave non-zero residual velocities distributed across the FoV; nevertheless, most of the characteristic patterns observed in the circular residuals are vanished. In both objects the $s_1$ harmonic is the non-circular term with the largest amplitude, { $\sim 0.2 c_1$}.

\subsection{V$_{max}$}
\label{subs:vmax}
Assuming that $V_t(r)$ and $c_1(r)$ approximately describe the circular rotation \citep{Franx1994,Wong2004},  we can estimate the maximum circular
velocity from the derived rotational curves.
Unlike $V_t$ from Eq.~\ref{Eq:circular}, $c_1$ takes into account the influence of mild non-circular motions on the disk; thus, we expect $c_1$ to trace a more accurate rotation curve than $V_t$.

We parameterize the rotation curves $V_t(r)$ and $c_1(r)$ adopting the following expression \citep[e.g.,][]{Courteau1997}:
\begin{equation}
 \label{Eq:Courteau}
 V_\mathrm{rot}(x) =  V_0\frac{(1+x)^\beta}{(1+x^\gamma)^{(1/\gamma)}}~~\mathrm{with~\it{x = r/r_t}},
\end{equation}
where $V_0$ is the asympthotic velocity, $r_t$ is the transition radius from rising to flat rotation and $\beta$ and $\gamma$ define the shape of the rotation curve; we assumed for all cases $\beta = 0$.

For each object in our kinematic sample, we estimate the maximum rotational velocity ($V\mathrm{_{max}}$) at $r_\mathrm{max} = 2.15h$ with $h$ being the disk scale. It has been observed that $V\mathrm{_{max}}$ computed at this radius reduces the most the scatter of the Tully-Fisher relation in optical rotation curves \citep[e.g.,][]{Courteau1997}.
In terms of the effective radius, and assuming the light is distributed as an exponential disk, we have  $r_e \sim 1.68h$ and $r_\mathrm{max} = 1.28r_e$.
However, it is well known that galaxies do not reach their maximum rotation within their optical extension. Thus, parameterizing them through Eq.~\ref{Eq:Courteau} can help us to estimate $V_{\rm max}$.

\subsection{Non parametric measurement of NC motions}
\subsubsection{Global strength: $V_{80}/V_{max}$}
The bottom panels in figure~\ref{fig:circular} showed the distribution of the residual velocities for our showcase objects. Although the median of the distributions centers around 0~\kms, indicating that a large fraction of spaxels are compatible with pure circular rotation, there is another fraction of spaxels whose velocities cannot be reproduced by adopting simple circular rotation models. The latter is true regardless of the considered velocity map. 
If noise-dominated spaxels are successfully removed from a velocity map, the tails of the residual distribution will provide information about non-circular motions and their sources. The narrower this distribution, the smaller the contribution of non-circular motions to the velocity map. For comparison, the distribution of residuals in our kinematic sample is shown in Figure~\ref{fig:hist_circ}. It is described by a Cauchy distribution centered in 0~\kms~with a 5~\kms~width. This figure reveals that circular rotation motions dominate in the majority of spaxels in our objects.

Since different sources can produce non-circular motions of different amplitudes, reducing a map of residual velocities (as in Fig.~\ref{fig:circular}) to a single characteristic value per object is not straightforward. Additionally, each galaxy is expected to have a unique residual profile.

Different non-parametric methods have been used in the literature to characterize the shape of a general distribution.
For instance the FWHM of emission lines, the width of the 80\% of the total flux of an emission-line to characterize outflow velocities \citep[e.g.,][]{Harrison2014}; or even the effective radius of a galaxy, which comprises 50\% of the total light. Often, these quantities do not a have a physical motivation, but they have been adopted systematically to characterize different properties of galaxies.
In this sense, the residual velocities obtained by subtracting the circular rotation from a velocity map represent the distribution of all sources of non-circular motions within a galaxy plus random noise and systematics.

With the aim of performing a homogeneous study of the non-circular motions in our sample, we adopt the width of the residual distribution that contains 80\% of the non-circular velocities. That is, $V_{80} = (V_{90}-V_{10})/\sin i$, where $V_{90}$ and $V_{10}$ represent the velocity at the 90th and 10th percentile of the cumulative distribution of the residual map (namely, observed-model), and $i$ is the disk inclination. We note that similar definitions have been adopted in the literature to characterize non-circular motions \citep[e.g.,][]{Erroz-Ferrer2015}. We correct $V_{80}$ by inclination, as other sources of non-circular motions, such as spiral arms or oval-distortions, lie on the disk plane. Since $V_{80}$ is estimated from the tails of the residual distribution, it tends to quantify the largest amplitudes from all different sources of non-circular motions.
For an axisymmetric velocity field, the distribution of residuals should be characterized by a Gaussian function centered around zero, with $V_{80} = 2.56 \,\sigma \sim$~FWHM (see appendix \ref{sec:appendix}).

In our kinematic sample, $V_{80}$ traces the average amplitude of the non-circular motions contained within an $1.5 r_e$ aperture, on average.
To compare the strengths of the non-circular motions among objects with different stellar mass,  we normalize $V_{80}$ obtained from circular rotation  by the characteristic maximum flat velocity obtained in Sec~\ref{subs:vmax}. As a consequence we define a new parameter $\eta$, as follows:
\begin{equation}
\eta = \log (V_{80}/V_\mathrm{max}).
\end{equation}
In this way, $\eta$ represents a characteristic strength of the non-circular motions in a galaxy, and simultaneously provides a systematic method for measuring non-circular motions in our sample.

Unlike other  methods to characterize non-circular motions, which often involve linear combinations of high harmonic terms normalized by the circular component $c_1$, ours does not depend on any non-circular model, but just on the residual velocities of a circular rotation model.

\subsubsection{Spatially resolved strength}
\label{subs:resolvnc}
 Another method to quantify the relative strength of NC motions in galaxies is by comparing the residual velocity map to the local circular rotation traced by the rotational curve.
Similar to $\eta$, we correct the residual maps by inclination and normalize them by $V_t(r)$, {\it i.e.}, the 2D rotational curve. This approach allow us to measure the amplitude of the non-circular motions relative to the local circular velocity. The normalization of residual velocities have been considered in previous works \citep[e.g.,][]{Andersen2013, Erroz-Ferrer2015}, although in this study, we apply a pixel-by-pixel normalization using the rotational curve.
This method allow us to spatially compare the relative strength of NC motions among galaxies with different stellar masses.

\begin{figure}
\centering
\includegraphics[]{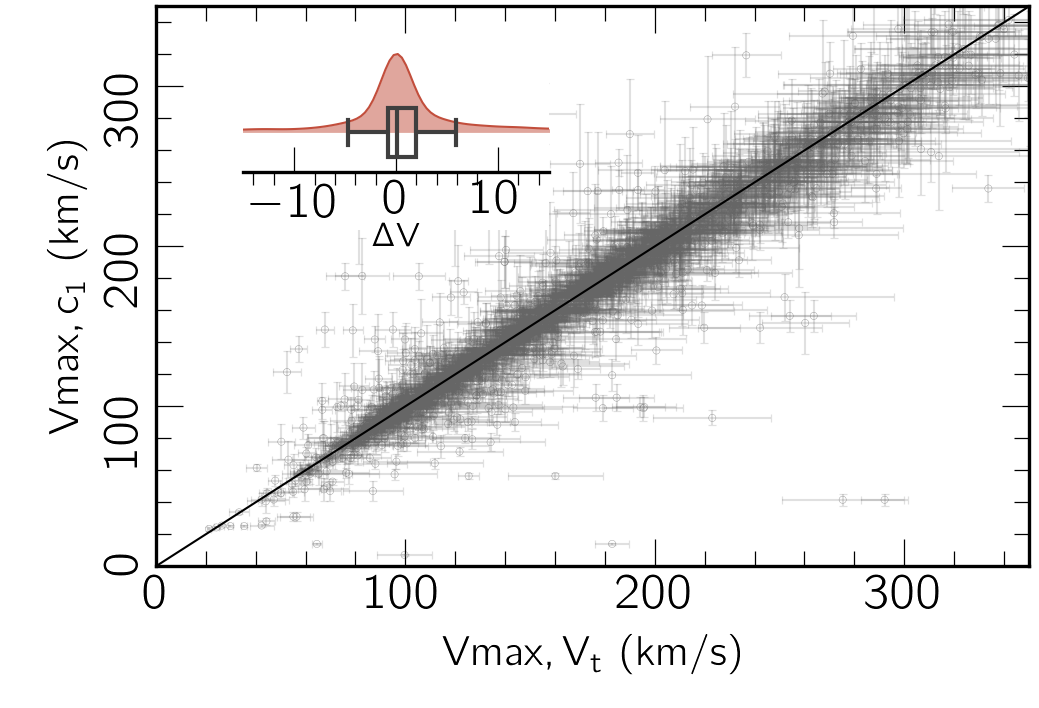}
\caption{Comparison between  maximum rotational velocities estimated from the circular rotation velocities $V_t(r)$, and the $c_1(r)$ velocities from the harmonic model. Every point here represents one individual object. The black straight line represents the 1-to-1 line. The inset panel shows a violin plot of the difference between both velocities, {\it i.e.}, $\Delta V = V_{max,c1} - V_{max,t} $. The dispersion of the difference is 5~\kms.}
\label{fig:c1_vs_vt}
\end{figure}

\section{Results}

\subsection{$V_t$ vs. $c_1$}

As noted from the showcase examples in figure~\ref{fig:harmonic}, the incorporation of non-circular terms in the harmonic model captures underlying non-circular motions that were previously unaccounted for in $V_t$. %This might might be crucial for the correct estimation of $V_{max}$.
Figure~\ref{fig:c1_vs_vt} shows the comparison of $V_\mathrm{max}$ in our sample adopting both kinematic models. It is evident that there is a good agreement between the estimates of  $V_\mathrm{max}$. Furthermore, when computing the difference in maximum speeds, $\Delta V_{max} = V_{max,c1} - V_{max,t}$, we observe that the scatter of this difference is minimal, only $5$~\kms.

Overall, we do not find significant differences when considering non-circular motions in computing $V_\mathrm{max}$. However, this does not mean that locally they are not important, as observed previously. Globally, non-circular motions do not appear to notably affect the characteristic rotation of galaxies, at least not in the optical rotation curves addressed here. This is somewhat expected for galaxies hosting bars, where most non-circular motions are confined to the bar region, { while $V_{max}$ typically occurs at larger radii, a few times the effective radius}.

\subsection{The Stellar Tully--Fisher relation including non-circular motions}

The stellar Tully Fisher relation (TFR) is a linear relation between the maximum rotation velocity of a galaxy, with its stellar mass  \citep[e.g.,][]{TullyFisher}. When the TFR is computed it is often assumed that non-circular motions are negligible, namely that residuals from circular rotation models represent imperfections in the modeling, but not sources of non-circular motions that affect to $V_{max}$.

Expressing the stellar mass as independent variable, the TFR adopts the following expression:
\begin{equation}
\label{Eq:TFR}
 \log V_{circ} = s[\log (M_{\star}/M_{\odot}) + x_0] + I,
\end{equation}
where $x_0$ is a term introduced to reduce the covariance between the slope ($s$) and intercept ($I$), usually estimated from the median value of the sample, in our case $x_0 = 10.5$. The value of the slope varies from 0.25-0.35 depending on the velocity definition adopted \citep[e.g.,][]{Hall2012, Ponomareva2018, Lelli2019, Stone2021}.
{ The scatter in the Tully Fisher relation has been attributed to the presence of non-circular motions caused by non-axisymmetric potentials \citep[e.g.,][]{Franx1992}. This suggests that non-circular motions have some impact in the inferred rotational curves.
To minimize the scatter in the TFR due to rising, falling, or wiggling rotation curves, we choose to include only flat rotational curves where we are confident in measuring the maximum rotation speed.} We achieve this by following the automated algorithm from \cite{Lelli2016}. Following this procedure, 454 objects were found to exhibit {\it bonafide} flat rotational curves. We then estimate  $V_{max}$ at $2.15h$ thorough linear interpolation of these rotational curves.

\begin{figure*}
\centering
\includegraphics[]{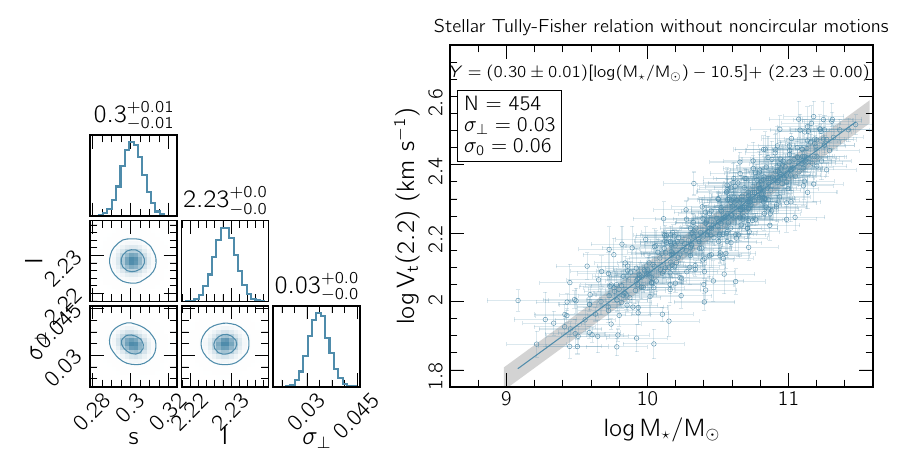}
\includegraphics[]{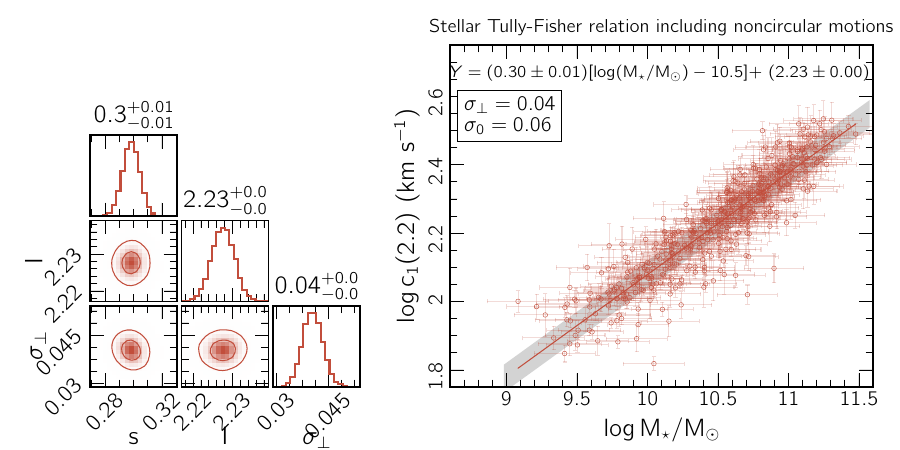}
\caption{Stellar Tully-Fisher relation without including (blue colors) and including (orange colors) non-circular motions for objects with flat rotational curves in our kinematic sample. The top-left panel shows the marginalized posterior distributions for the slope $(s)$, zero point $(I)$, and perpendicular scatter $(\sigma_{\perp})$ of the TFR without considering non-circular motions. Median values and $\pm 1\sigma$ dispersion for each parameter are shown at the top of the histograms. The top-right panel shows the stellar TFR with the best-fit parameters and intrinsic dispersion shown in gray color. $V_{max}$ is estimated from $V_t(r)$  at $2.2h$. The bottom-left and bottom-right panels have similar meanings as above but include non-circular motions via harmonic decomposition of the LoS velocities. In this case $V_{max}$ is estimated with $c_1(r)$ also at $2.2h$.
}
\label{fig:TF}
\end{figure*}

We performed an orthogonal distance regression (ODR) with intrinsic scatter to estimate the slope and intercept. This method takes into account the error on the $XY$ variables, namely  the stellar mass and $V_{max}$ respectively, while minimizing the perpendicular distance from the best fit line. To obtain the best estimation of the uncertainties on the slope and intercept, we adopt Bayesian methods. Specifically, we implement dynamic nested sampling \citep[NS,][]{NS, DNS} via {\sc dynesty } \citep[e.g.,][]{Speagle2020}, which is a Python implementation of the dynamic NS algorithm.
The likelihood function adopted is similar to that of \citet{Lelli2019}, but considering the stellar mass and maximum velocity as the independent and dependent variables, respectively,  following Equation~\ref{Eq:TFR}.  The results of this analysis is shown in Figure~\ref{fig:TF}.

Assuming pure circular rotation ({\it i.e.}, Eq.~\ref{Eq:circular}) in the computation of $V_{max}$ we find a slope of $0.30\pm0.01$ dex, with an intrinsic orthogonal scatter of 0.03 dex and a dispersion of 0.06 dex. This slope value is in total agreement with other results from MaNGA considering only circular motions \cite[e.g.,][]{Erik2020}, although differs by approximately $0.02$ dex from recent results from \citet{Arora2023}, most likely due to differences in the modeling of the rotation curve and the definition of $V_{max}$.
On the other hand, when we include non-circular motions in the estimation of $V_{max}$ ({\it i.e.}, Eq.~\ref{Eq:m=3}) we find that the TFR slope is $0.30\pm0.01$, while the intrinsic scatter remains unchanged. As observed, neither the slope nor the global scatter in the TFR are affected when considering the presence of non-circular motions in the estimation of $V_{max}$.
These results may suggest that non-circular motions are not responsible of the observed scatter in the TFR, or at least this is not observed in optical rotational curves.

\subsection{MaNGA radial profiles of NC motions}
\label{sec:MaNGA_radial_profiles}

\begin{figure}
\centering
\includegraphics[]{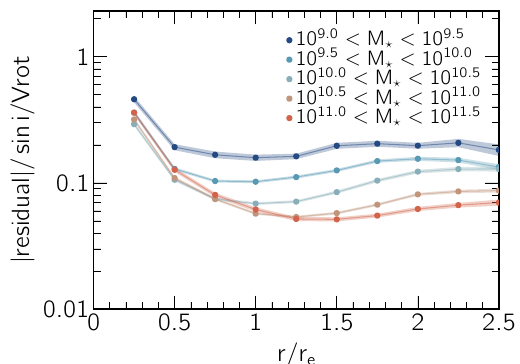}
\caption{ Radial distribution of the circular model residuals (in absolute value) normalized by the rotational curve $V_t(r)$ for the 1624 objects in our kinematic sample. Each dot represents the median value of the residuals in $0.25r_e$ bins. Galaxies are segregated into 5 different mass bins represented by different colors.}
\label{fig:radial_profiles1}
\end{figure}

\begin{figure}
\centering
\includegraphics[]{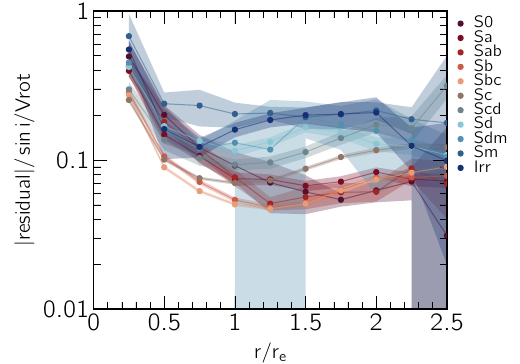}
\caption{Similar figure as fig.~\ref{fig:radial_profiles1}, but this time galaxies have been segregated by morphological types from S0 to Irregulars.
}
\label{fig:radial_profiles2}
\end{figure}

{ Deviations from axisymmetry in galaxies are induced by non-axisymmetric potentials; consequently, it is expected that the amplitude of non-circular motions induced by the non-axisymmetric structures, such as bars, bulge, spiral arms, is also related to the stellar mass.}
With the normalized residual maps described in section~\ref{subs:resolvnc}, we built radial profiles of the residual velocities (in absolute value and corrected by inclination to take into account sources of non-circular motions on the disk plane) for the objects in our kinematic sample. For this purpose we deproject the gaseous disks adopting the inclination and position angles derived from our kinematic analysis. We create radial bins of $0.25r_{e}$ widths and group galaxies by stellar mass and morphological type as observed in Figures~\ref{fig:radial_profiles1} and \ref{fig:radial_profiles2}, respectively.

In the former figure, we find that the radial profiles are clearly separated depending on the stellar mass bin, following
almost flat distributions. { This behavior is probably a consequence of the normalization of the residuals by the rotation curve, where the residuals are roughly constant with mass.}
{ In the innermost radii, below $0.75r_{e}$, we observe a sharp increase ($\sim\times4$) in the non-circular motions with respect to the circular rotation. This increase becomes more pronounced at higher mass where the rise in the inner rotation curve becomes steeper. Also, the first radial bin encloses the bulge region in our sample, which is expected to be dominated by random motions.}

Additionally, from this figure, we observe that the amplitude of the non-circular motions in massive galaxies can be as large as 5\% of the local circular rotation, while this ratio of non-circular to circular amplitudes can reach up to 20\% in less massive galaxies.

The distribution by morphology is shown in Figure~\ref{fig:radial_profiles2}. It shows a similar behavior as the stellar mass,
although the lack of morphological information for all galaxies is clearly affecting the mean values as reflected by the error bars.
{ The morphology distribution shows that the latest type spirals (Sd) have  larger amplitudes of NC motions, $\sim4\times$, compared to early type spirals (Sa).}

%At each galactocentric distance, the latest type spirals (Sd) tend to exhibit a larger ratio of non-circular to circular motions, while this ratio is lower for early-type spirals (Sa).

\subsection{The $\eta$ parameter and global properties}
Unlike the spatially resolved normalization discussed in the previous section, $\eta$ is a global parameter that aims to measure the characteristic strength of non-circular rotation.
We investigate whether the maximum strength of non-circular motions in our kinematic sample is related to global properties.

\begin{figure}
\centering\includegraphics[]{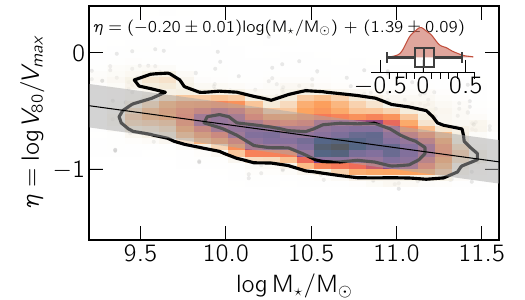}
\centering\includegraphics[]{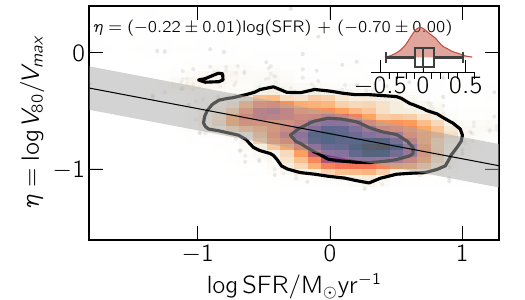}
\centering\includegraphics[]{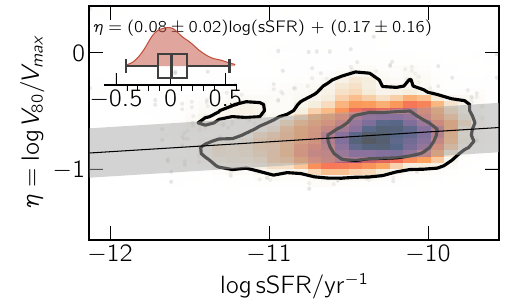}
\caption{ {\it Top panel:} 2D histogram showing the distribution of the $\eta$ parameter versus the stellar mass for the objects in the kinematic sample. Colors represent the density of points within each bin, with darker colors representing regions with a larger density of objects. The inner and outer contours enclose 68\% and 95\% percent of the data, respectively.
The black straight line represents the best Bayesian linear fit to these points, with $1\sigma$ scatter shown with gray color; the half violin plot shows the residuals respect the best fit line. {\it Middle and bottom panels:} Similar figures as above but for the integrated \ha~based SFR and specific SFR, respectively.}
\label{fig:q95Mstar}
\end{figure}

\subsubsection{Stellar mass, SFR and sSFR}
Similar to the circular rotation case, we examine whether the maximum non-circular velocities have a linear dependence on the stellar mass.
Figure~\ref{fig:q95Mstar} shows the relation between  $\eta$ and the stellar mass for our kinematic sample. This figure suggests that the global strength of non-circular motions in galaxies is indeed related with the stellar mass.
As in the TFR, we fit the best line adopting the form $\eta = m\log M_\star/M_\odot~ +~ b$, with $m$ and $b$ being the slope and intercept, respectively. We find a negative slope for this relation, which is $-0.20$.
This plot suggests that the effects of non-circular motions in galaxies can be different depending on the stellar mass.

The middle panel of Figure~\ref{fig:q95Mstar} shows the relation between $\eta$ and the global \ha~based star formation rate (SFR), after correcting the \ha~flux by dust attenuation \citep[e.g.,][]{Cardelli1989}, assuming case B of recombination \citep[e.g.,][]{osterbrock89} and redshift-based angular distances. This relation shows a slope of $-0.22$, similar as the previous one found with the stellar mass. Both the stellar mass and SFR exhibit moderate correlations in terms of the Spearman coefficient, ranging between $\rho = 0.4-0.5$.

Finally we compare $\eta$ with the specific SFR (sSFR = SFR/M$_{\star}$), this is shown in the bottom panel of Figure~\ref{fig:q95Mstar}. In this case we do not find a clear relation between these two parameters, showing a $0.08$ slope.

\section{Conclusions}
 
We investigated the role of non-circular motions in galaxies at both local and global scales using a carefully selected sample of 1624 star-forming galaxies from the MaNGA galaxy survey and adopting the MPL-11 products. 
Non-circular motions were analyzed using residual velocity maps derived from circular rotation models, a global parameter $\eta$ designed to quantify the characteristic strength of non-circular, and by modeling  the harmonic coefficients expected from a second order perturbation in the potential.

{ We found that the rotational curve traced with circular ($V_t$) and non-circular rotation models ($c_1$) show minimal differences, even though the harmonic model captures underlying non-axisymmetric motions. Consequently,
the maximum rotational velocity $V_{max}$ remains unaffected when considering non-circular motions. This study, therefore, indicates that non-circular rotation has minimal influence on the estimation of the
rotation speed of galaxies. Scaling relations involving $V_{max}$, such as the barionic Tully-Fisher relation, show no changes in the slope, zero point, or scatter when non-circular motions are considered.
However, this conclusion may be influenced by the assumed non-circular rotation model, which is intended to account for small perturbations.
If confirmed with larger samples and kinematic models that account for arbitrary amplitudes of NC motions, this would suggest that non-circular motions do not contribute to the intrinsic scatter in the TFR.

Residual velocity maps, while useful for for characterizing the local and global strength of non-circular motions in galaxies, however the choice of the best normalization method for describing these motions remains unclear.
Interestingly, both local and global measurements of NC motions reveal a correlation between the average amplitudes of NC motions and the stellar mass. However, large and more homogeneous samples, with a wide covering  of stellar masses and improved spatial resolution, are necessary to better understand and refine the observed relationship.
}

\section*{acknowledgments}
CLC acknowledges support provided by Academia Sinica Institute of Astronomy and Astrophysics.
LL acknowledges the Ministry of Science \& Technology of Taiwan for providing support through the grants NSTC 112-2112-M-001-062- and NSTC 113-2112-M-001-006-.
JN acknowledges funding from the European Research Council (ERC) under the European Union’s Horizon 2020 research and innovation programme (grant agreement No. 694343).
This project makes use of the MaNGA-Pipe3D dataproducts. We thank the IA-UNAM MaNGA team for creating this catalogue, and the Conacyt Project CB-285080 for supporting them.

Funding for the Sloan Digital Sky
Survey IV has been provided by the
Alfred P. Sloan Foundation, the U.S.
Department of Energy Office of
Science, and the Participating
Institutions.

SDSS-IV acknowledges support and
resources from the Center for High
Performance Computing  at the
University of Utah. The SDSS
website is www.sdss4.org.

SDSS-IV is managed by the
Astrophysical Research Consortium
for the Participating Institutions
of the SDSS Collaboration including
the Brazilian Participation Group,
the Carnegie Institution for Science,
Carnegie Mellon University, Center for
Astrophysics | Harvard \&
Smithsonian, the Chilean Participation
Group, the French Participation Group,
Instituto de Astrof\'isica de
Canarias, The Johns Hopkins
University, Kavli Institute for the
Physics and Mathematics of the
Universe (IPMU) / University of
Tokyo, the Korean Participation Group,
Lawrence Berkeley National Laboratory,
Leibniz Institut f\"ur Astrophysik
Potsdam (AIP),  Max-Planck-Institut
f\"ur Astronomie (MPIA Heidelberg),
Max-Planck-Institut f\"ur
Astrophysik (MPA Garching),
Max-Planck-Institut f\"ur
Extraterrestrische Physik (MPE),
National Astronomical Observatories of
China, New Mexico State University,
New York University, University of
Notre Dame, Observat\'ario
Nacional / MCTI, The Ohio State
University, Pennsylvania State
University, Shanghai
Astronomical Observatory, United
Kingdom Participation Group,
Universidad Nacional Aut\'onoma
de M\'exico, University of Arizona,
University of Colorado Boulder,
University of Oxford, University of
Portsmouth, University of Utah,
University of Virginia, University
of Washington, University of
Wisconsin, Vanderbilt University,
and Yale University.

%\section{Conclusions}
\newpage
\appendix
\section{Additional figures}
\label{sec:appendix}
%Figure~\ref{fig:bisymmetric} shows the bisymmetric model from \xs~for the barred galaxy shown in Figure~\ref{fig:circular}. As the harmonic model, this one also provides a physical motivated description for the observed non-circular motions in this object. In this parameterization two velocities describe the non-circular motions, the tangential $V_{2t}$ and the radial $V_{2r}$, both originated because of the bar. However, this model cannot be applied systematically to all our sample since it assumes non-circular motions are preferentially aligned towards a constant position angle, namely $\phi_{bar}$. Thus it does not have any meaning for spiral galaxies without a bar.

\begin{figure}
\centering
\includegraphics[width = 0.57\textwidth]{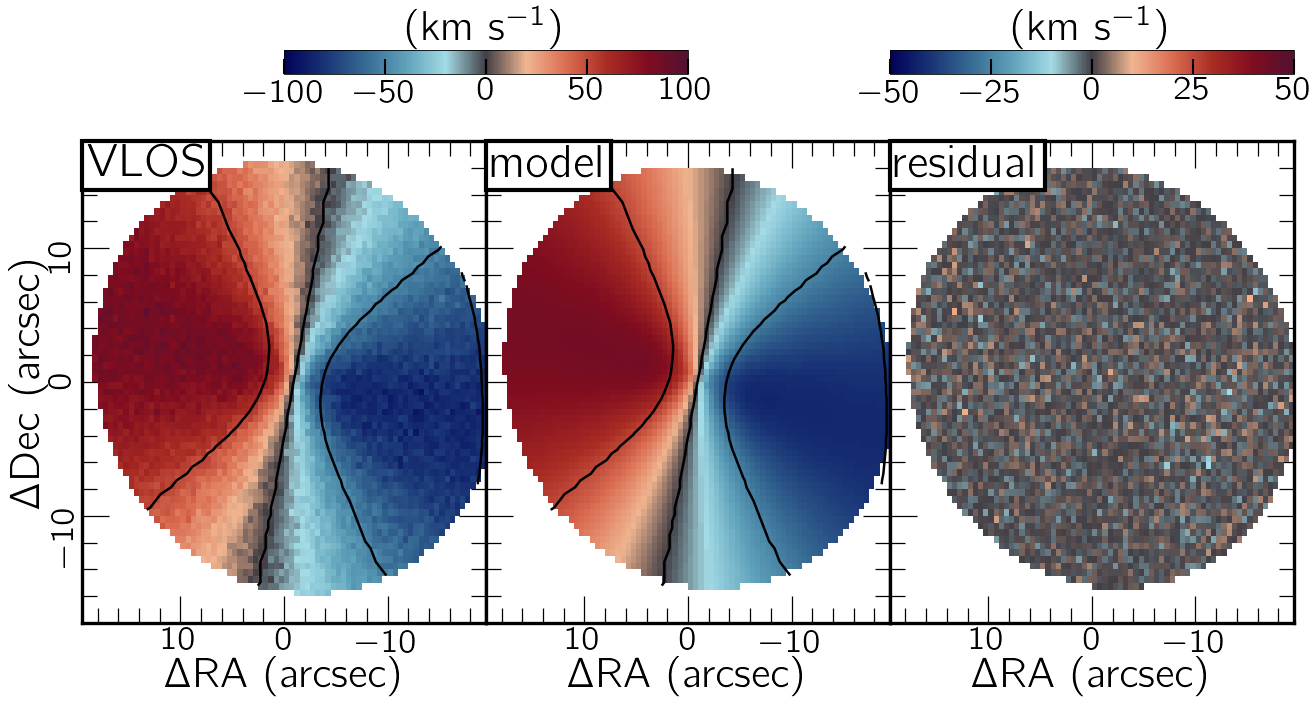}
\includegraphics[width = 0.42\textwidth]{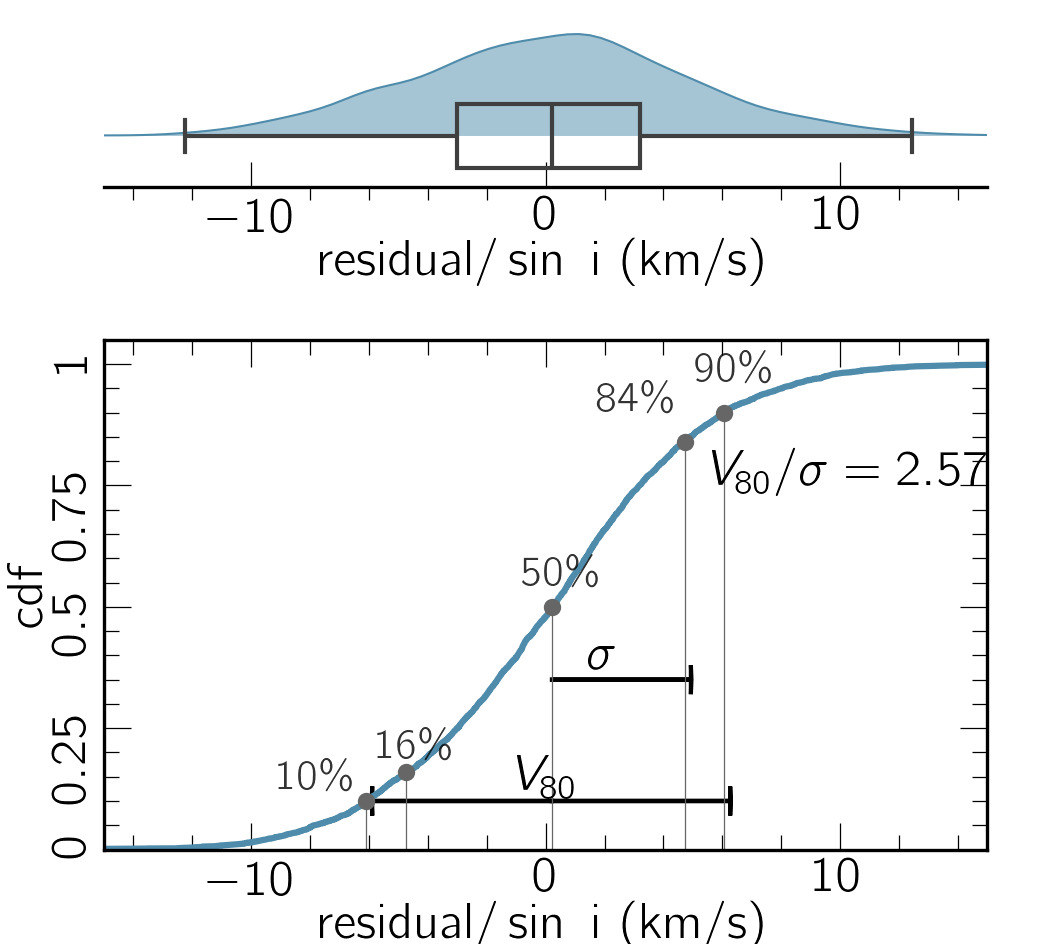}
\caption{{\it First panel:} Simulated axisymmetric velocity field plus random noise; the is inclined at $i=35^{\circ}$ and is oriented at $\phi_{disk}^{\prime}=77^{\circ}$, the rotation curve adopted follows Eq.~\ref{Eq:Courteau} with $V_0 = 170$~\kms, and $r_t = 2\arcsec$. {\it Middle panel:} Model recovered using \xs. {\it Third panel:} Residual map, namely, observed minus model. {\it Fourth panel:} cumulative distribution of the residual velocities. Different percentiles are highlighted, going from 10\% ($V_{10}$) to 90\%($V_{90}$).  $1\sigma$ encloses 84\% of the data, while $V_{80}= (V_{90}-V_{10})/\sin i$. The ratio $V_{80}/\sigma=2.67$ which is approximately to 1.1~FWHM.}
\label{fig:circ_gaussian}
\end{figure}

Figure~\ref{fig:circ_gaussian} shows $V_{80}$ and $\sigma$ for an axisymmetric velocity field. It follows that $V_{80}=1.1$\,FWHM.

%Figure~\ref{fig:outflow} shows a clear example of a galaxy in which non-circular motions are not driven by non-axisymmetric potentials. Instead, a SF-driven bipolar outflow is disturbing the ionized gas kinematics, affecting the LoS-velocities. The harmonic model from Eq.~\ref{Eq:m=3} does not provide a physical interpretation for the non-circular velocities $s_1$, $s_3$, $c_3$.

%\begin{figure*}
%\centering
%\includegraphics[]{kin_bisymmetric_model_manga-10842-12704-LM.png}
%\caption{Bisymmetric model  describing the oval distortion observed in the barred galaxy manga-10842-12704 also shown in Figure~\ref{fig:circular}. The best fit parameters from this model are: $\phi_{disk}^{\prime} = 353^{\circ},~i = 61^{\circ}$, Vsys = $8483$~\kms; the oval distortion is orientated at $\phi_{bar}=143^{\circ}$ or $333^{\circ}$ projected in the sky. The latter angle roughly coincides with the visual photometric position angle of the bar. This model is exclusive for bar-like motions, therefore it can not be applied in a systematic way in all our sample. }
%\label{fig:bisymmetric}
%\end{figure*}

%\begin{figure*}
%\centering
%\includegraphics[]{kin_circular_model_manga-8990-9101.png}
%\caption{MaNGA galaxy 8990-9101 exhibiting a strong bipolar outflow induced by star-formation. {\it From left to right:} (i) continuum image reconstructed from the MaNGA $gri$ bands; (ii) false color image showing the distribution of ionized gas traced by \nii~(red), \oiii~(blue) and \ha~(green); (iii) \ha~velocity map; (iv) best circular rotation model from \xs; (v) residual map from the model, namely \ha~vel.~map-model; (vi) rotational curve.
%
%The \ha~velocity field is clearly affected by the outflow observed in reddish colors in the emission-line image. The residuals velocities from a circular rotation model show several patterns associated to the expanding cones. A lower \sn~cut in \ha~has been applied to highlight the outflow.}
%\label{fig:outflow}
%\end{figure*}


\bibliography{tesis}{}
\bibliographystyle{aasjournal}



\end{document}